\documentstyle[preprint,aps,prb]{revtex}
\begin{document}
\title{
Coulomb-correlation effects in semiconductor quantum dots:  \\
The role of dimensionality 
}
\author{Massimo Rontani, Fausto Rossi, Franca Manghi, and Elisa Molinari}
\address{
Istituto Nazionale per la Fisica della Materia (INFM), and \\
Dipartimento di Fisica, Universit\`a di Modena, \\
via Campi 213/A, I-41100 Modena, Italy
}
\date{\today}
\maketitle
\begin{abstract}
We study the energy spectra of small three-dimensional (3D)
and two-dimensional (2D) semiconductor quantum dots through 
different theoretical approaches
(single-site Hubbard and Hartree-Fock hamiltonians);
in the smallest dots we also compare with exact results.
We find that purely 2D models often lead
to an inadequate description of the Coulomb
interaction existing in realistic structures, 
as a consequence of the overestimated carrier localization. 
We show that the dimensionality of the dots has a crucial impact 
on (i) the accuracy of the predicted addition spectra; (ii) the
range of validity of approximate theoretical schemes. 
When applied to realistic 3D geometries, the latter are found
to be much more accurate than in the corresponding 2D cases
for a large class of quantum dots; the single-site Hubbard
hamiltonian is shown to provide a very effective and accurate 
scheme to describe quantum dot spectra, leading
to good agreement with experiments.

\end{abstract}
\clearpage
\narrowtext

\section{Introduction}

Adding an electron into a semiconductor quantum dot (QD) produces a
variation in the energy of the system that depends
on single-particle quantum confinement as well as on the Coulomb interaction
between carriers.\cite{rev_exp}
Understanding such {\it addition-energy spectrum} is a key step 
towards controlling the
physics of single-electron devices. At the same time, the addition spectra
of quantum dots offer a unique probe of few-particle interactions in
regimes that are not experimentally accessible in atomic physics. The
experimental effort in this direction has developed very rapidly after the
recent fabrication of controlled small-QD devices based on gated vertical 
heterostructures\cite{Tarucha} or self-assembled dots.\cite{Fricke}
The resulting addition spectra
show a clear shell structure, corresponding to the symmetries of the
confining potential, with a filling sequence analogous to Hund's rule in
atomic physics.

From the theoretical point of view, a general interpretation of these
features was obtained by calculating the energy spectrum for a strictly
two-dimensional (2D) quantum dot, and using either exact methods (for very
few electrons), or approximate ---usually Hartree-Fock--- 
methods.\cite{rev_th} 
The assumption of a purely 2D model was initially motivated by the
typical disk-like shape of the QD potential, whose extension along $z$ is
(slightly) smaller than the lateral extension of the carrier ground state
in the $xy$ plane. If one adopts a separable picture for the QD confining
potential, $V=V\!\left(z\right)+V\!\left(x,y\right)$,
the relevant (i.e. lowest) single-electron states can be all
associated to the ground state of $V\!\left(z\right)$.
From the point of view of single-particle states the 2D assumption is 
therefore justified.

In view of the three-dimensional (3D) nature of the Coulomb interaction,
however, the 2D model introduces additional approximations in the
calculation of the Coulomb integrals, which are sensitive to the spatial
extension ---2D vs. 3D--- of the single-particle wave 
functions.\cite{noi,threedim} 
In turn, Coulomb integrals control electron-electron correlation, and
influence the quantitative determination of addition spectra and their
dependence on magnetic field. At the same time, the strength of Coulomb
interaction is also the key parameter determining the accuracy and range of
validity of the approximations which must be introduced for dots with many
electrons.

In this paper we investigate theoretically the addition spectra of 
realistic QD structures, with special
emphasis on the effects of electron-electron repulsion and their dependence
on the geometry and dimensionality of the confining potential. 
In Sect.~\ref{general}, we compare different
approximate solutions of the general Hamiltonian
for $N$ interacting electrons confined in a QD structure; in particular
we consider the single-site Hubbard (SSH) scheme introduced in
Ref.~\onlinecite{noi}, and the standard Hartree-Fock (HF) method.

In Sect.~\ref{elio}, we focus on the simplest
case, i.e., a two-electron system within
a parabolic confining potential, and calculate the exact energy eigenvalues
and pair correlation functions for the 2D and 3D case. 
As in Ref.~\onlinecite{Pfannkuche},
we use this prototypical system
---called artificial or QD Helium--- as a reference to evaluate 
the accuracy of the different approximation schemes: 
We find that both the importance of corrections
beyond HF and the differences between HF and SSH are drastically reduced
for a realistic 3D description of the dot with respect to its 2D
modelization, mainly as a consequence of the reduced Coulomb integrals. 
This suggests the reliability of a fully 3D mean field treatment of
semiconductor QDs.

Section \ref{results} is then devoted to
the application of HF and SSH methods to 3D
and 2D quantum dots with a larger number of electrons. We compare both
methods for QD structures of different geometries and demonstrate that SSH is an 
accurate and efficient scheme for realistic, i.e. 3D-like, dots.
Finally, we discuss the implications of our results for the interpretation of 
recent experimental data vs magnetic field in QD structures and draw 
some conclusions.

\section{Theoretical approach: Exact formulation and approximation schemes}
\label{general}

Our aim is to describe $N$ electrons, confined in a QD structure (with harmonic 
in-plane confining potential) and interacting via Coulomb law,
possibly in the presence of an external
magnetic field perpendicular to the plane.
The general $N$-particle Hamiltonian is
\begin{equation}
\hat{\cal{H}} = \sum_{i=1}^N \hat{H}_0\!\left(i\right)
+\frac{1}{2}\sum_{i\neq j}
\frac{e^2}{\kappa \left| \bbox{r_i}-\bbox{r_j} \right|} ,
\label{eq:hmanybody}
\end{equation}
where the single particle Hamiltonian, within the effective mass approximation,
is
\begin{equation}
\hat{H}_0\!\left(i\right)
=\frac{1}{2m^{\ast}}\left(\bbox{\hat{p}}+
\frac{e}{c}\bbox{\hat{A}}\left(\bbox{r}_i\right)\right)^2
+\frac{1}{2}m^{\ast}\omega^2_0\!\left(x_i^2+y_i^2\right)
+V\!\left(z_i\right) \ .
\label{eq:hgen}
\end{equation}
Here $\bbox{\hat{A}}$ is the vector potential,
$\kappa$ and $m^{\ast}$ are the scalar dielectric costant and
the effective electron mass in the semiconductor, $\omega_0$ is the
characteristic oscillator frequency of the in-plane confining potential,
and $V\!\left(z\right)$
is the confining potential along $z$; $V\!\left(z\right)$
can be chosen either as a harmonic potential
($V\!\left(z\right)=\frac{1}{2}m^{\ast}\omega^2_0z^2$),
a square well or a zero-width
infinite barrier to describe  spherical, cylindrical or disk-shaped QD 
structures, respectively. 
Here Zeeman coupling between spin and magnetic field has been neglected.

This general  Hamiltonian can be written in second quantized form on the
complete and orthonormalized basis of single particle states
\begin{equation}
\hat{\cal{H}}=\sum_{\alpha \sigma}
\varepsilon_{\alpha}\hat{c}_{\alpha \sigma}^{\dagger}
\hat{c}_{\alpha \sigma}
+\frac{1}{2}
\sum_{\alpha\beta\gamma\delta}
\sum_{\sigma \sigma^{\prime}}
V_{\alpha\sigma,\beta\sigma^{\prime};\gamma\sigma^{\prime},\delta\sigma}
\hat{c}_{\alpha\sigma}^{\dagger}
\hat{c}_{\beta\sigma{\prime}}^{\dagger}
\hat{c}_{\gamma\sigma{\prime}}
\hat{c}_{\delta\sigma}.
\end{equation}
Here $\varepsilon_{\alpha}$  are the eigenenergies of the one-particle
Hamiltonian $\hat{H}_0$, $\hat{c}_{\alpha \sigma}^{\dagger}$ and
$\hat{c}_{\alpha\sigma}$ the creation and destruction operators for an
electron with orbital index $\alpha$ and spin $\sigma$;
$V_{\alpha\sigma,\beta\sigma^{\prime};\gamma\sigma^{\prime},\delta\sigma}$
are the two-body matrix elements of the electron-electron interaction
\[
V_{\alpha\sigma,\beta\sigma^{\prime};\gamma\sigma^{\prime},\delta\sigma}=
\sum_{s s'} \int\!\! \phi_{\alpha\sigma}^*\!\left(\bbox{r},s\right)
\phi_{\beta\sigma^{\prime}}^*\!\left(\bbox{r}^{\prime},s^{\prime}\right)
\frac{e^2}{|\bbox{r}-\bbox{r^{\prime}}|}
\phi_{\gamma\sigma^{\prime}}\!\left(\bbox{r}^{\prime},s^{\prime}\right)
\phi_{\delta\sigma}\!\left(\bbox{r},s\right)
{\rm\,d}\bbox{r}{\rm\,d}\bbox{{r}^{\prime}}
\]
where $\phi_{\alpha\sigma}
\!\left(\bbox{r},s\right)=\phi_{\alpha}\!\left(\bbox{r}
\right)\chi_{\sigma}\!\left(s\right)$
are the single particle eigenfunctions.

It is useful to isolate among the Coulomb matrix elements the
``semi-diagonal'' ones, namely
\[
V_{\alpha\sigma,\beta\sigma;\beta\sigma,\alpha\sigma}=
V_{\alpha\sigma,\beta -\sigma;\beta -\sigma,\alpha\sigma}
\equiv
U_{\alpha \beta}
\]
\[
V_{\alpha\sigma,\beta\sigma;\alpha\sigma,\beta\sigma}
\equiv J_{\alpha \beta}.
\]
These are the usual direct and exchange integrals which can be written
more explicitely as
\begin{equation}
U_{\alpha\beta} = e^2\int\!\!\!\int
\frac{ {\left|{\phi}_{\alpha}\!\left(\bbox{r}\right)\right|}^2
{\left|{\phi}_{\beta}\!\left(\bbox{r^{\prime}}\right)\right|}^2}
{\kappa\left|\bbox{r}-\bbox{r^{\prime}}\right|}
{\rm\,d}\bbox{r}{\rm\,d}\bbox{r^{\prime}}
\label{eq:Coulomb}
\end{equation}
\begin{equation}
J_{\alpha\beta} = e^2\int\!\!\!\int
\frac{ {{\phi}_{\alpha}}^{\!\ast}\!\left( \bbox{r} \right)
{{\phi}_{\beta}}^{\!\ast}\!\left(\bbox{r^{\prime}}\right)
{\phi}_{\alpha}\!\left(\bbox{r^{\prime}}\right)
{\phi}_{\beta}\!\left(\bbox{r}\right) }{\kappa\left|\bbox{r}-\bbox{r^{\prime}}
\right|}
{\rm\,d}\bbox{r}{\rm\,d}\bbox{r^{\prime}}.
\label{eq:exchange}
\end{equation}
In this way eq.~(\ref{eq:hgen}) becomes
\begin{equation}
\hat{\cal{H}} =
\hat{\cal{H}}^{SSH}
+
\frac{1}{2}{\sum_{\alpha\beta\gamma\delta}}^{\prime}
\sum_{\sigma\sigma^{\prime}}
V_{\alpha\sigma,\beta\sigma^{\prime};\gamma\sigma^{\prime},\delta\sigma}
\hat{c}_{\alpha\sigma}^{\dagger}
\hat{c}_{\beta\sigma{\prime}}^{\dagger}
\hat{c}_{\gamma\sigma{\prime}}
\hat{c}_{\delta\sigma};
\label{eq:hh}
\end{equation}
where the prime on the first summation is to omit the terms with
$\alpha=\delta$, $\beta=\gamma$ and $\alpha=\gamma$,
$\beta=\delta$ and
\begin{equation}
\hat{\cal{H}}^{SSH} =
\sum_{\alpha\sigma}\varepsilon_{\alpha}\hat{n}_{\alpha\sigma}+
\frac{1}{2}\sum_{\alpha\beta\sigma}
\left[\left(U_{\alpha\beta}-J_{\alpha\beta} \right)
\hat{n}_{\alpha\sigma}\hat{n}_{\beta\sigma} +
U_{\alpha\beta}\hat{n}_{\alpha \sigma}\hat{n}_{\beta -\sigma}\right].
\label{eq:SSH}
\end{equation}
The relevance of this formal partition is twofold: i) it naturally leads to
a perturbation expansion in the off-diagonal interactions which are in
general smaller than the semi-diagonal ones; ii) moreover, the {\it
unperturbed} term $\hat{\cal{H}}^{SSH}$ is one-body like, with single Slater
determinants as {\it exact} eigenstates. The SSH approach defined in
Ref.\onlinecite{noi} consists in assuming $\hat{\cal{H}} \simeq
\hat{\cal{H}}^{SSH}$, which amounts to neglecting the second and 
higher order
contributions in the off-diagonal interactions, the first order one being
exactly zero.

The assumption that the off-diagonal $V_{\alpha,\beta;\gamma,\delta}$ are
negligible with respect to the semi-diagonal ones is implicit in all the
methods which describe electron correlation in terms of the Hubbard model,
either in its original form,\cite{Hubbard}
including only on-site interaction between
opposite spin electrons ---proportional to $U_{\alpha \beta}$---,
or adding the interaction
between parallel spin electrons as well, proportional to $(U_{\alpha \beta}
-J_{\alpha \beta})$. 
The important point here is that when the Hubbard model is applied 
to an isolated  QD,
i.e.~to a single site, the Hubbard Hamiltonian turns out to be one-particle
like: this is so because the inter-site hopping of the traditional Hubbard
Hamiltonian is absent in this case and the commutator
$\left[\hat{\cal{H}}^{SSH}, \hat{n}_{\alpha\sigma}
\right]$ is zero. As a consequence, the
Slater determinants, eigenstates of the single particle Hamiltonian
$\hat{H}_0$,  are  exact eigenstates of
$\hat{\cal{H}}^{SSH}$ as well.

Within the SSH approach the total energy of $N$ electrons in a QD 
structure is given by
\begin{equation}
E^{SSH}\!\left(N\right)=
\left<\Phi^N\right|\hat{\cal{H}}^{SSH}\left|\Phi^N\right> =
\sum_{\alpha \sigma} \varepsilon_{\alpha} \langle \hat{n}_{\alpha\sigma}
\rangle + \frac{1}{2}
      \sum_{\alpha\beta \sigma} \left[U_{\alpha\beta}
\langle \hat{n}_{\beta -\sigma} \rangle +
\left(U_{\alpha\beta}-J_{\alpha\beta}\right)\langle\hat{n}_{\beta \sigma}
\rangle\right]
      \langle \hat{n}_{\alpha \sigma} \rangle,
\label{eq:E}
\end{equation}
where $\left|\Phi^N\right>$ is a Slater determinant, eigenvector
of $\hat{H}_0$, and
$\langle\ \rangle$ denotes the average over the many-particle
eigenstate, which in our case simply reduces to the orbital
occupation number.

The proposed SSH approach shares in common with Hartree-Fock
methods the form of the total energy, which in both
schemes is expressed as the average of the exact Hamiltonian over 
a single Slater determinant;  
the variational prescription ---allowing for the construction of
optimal single-particle orbitals through the selfconsistent solution of a
single-particle eigenvalue problem--- is not present in the SSH approach. We
notice however that the importance of self-consistency is strongly related
to the relative weight of Coulomb matrix elements:
the HF potential entering the
self-consistent HF one-particle Hamiltonian is in fact related to the
direct and exchange Coulomb integrals; similarly the SSH approximation is
exact ---without any need of self-consistency--- whenever the higher-order
contributions from the off-diagonal Coulomb matrix elements is negligible. For
this reason we expect that a lower localization of the confined 
single-particle states in 3D with respect to 2D,
giving rise to smaller non-diagonal
Coulomb integrals, will
reduce the difference between HF and SSH results.
To check in detail this
point we have explicitely performed
HF calculations; we have used in particular
the matrix form of the unrestricted HF equation.\cite{Roothaan}

Whenever possible, it is obviously
useful to compare the outcomes of different approximate schemes
with exact results. This is done in the next section where we consider the
exactly solvable two-electron QD (artificial Helium)
in different confinement regimes; we will
show that the differences between HF and SSH results will be always
comparable with those between HF and exact results and that they scale with
the dimensionality of the confining potential.

\section{The two-electron problem}\label{elio}

In this section we will study the motion of two
electrons within a QD structure in two and three dimensions.
In this case, the exact Hamiltonian (\ref{eq:hmanybody}) reduces to
\begin{equation}
\hat{\cal{H}}=\hat{H}_0\!\left(1\right)+\hat{H}_0\!\left(2\right)
+
\frac{e^2}{\kappa\left|\bbox{r_1}-\bbox{r_2}\right|} ,
\label{eq:h2e}
\end{equation}
Here $\bbox{r}_i$ is the position of the electron,
$\bbox{r}_i\equiv\left(x_i,y_i\right)$ in 2D or
$\bbox{r}_i\equiv\left(x_i,y_i,z_i\right)$  in 3D,
and $\bbox{p}_i$ the corresponding momentum.

To solve this equation, we perform the standard transformation\cite{Merkt}
to center of mass (CM) coordinates,
$\bbox{R}=\left(\bbox{r}_1+\bbox{r}_2\right)/2$, $\hat{\bbox{P}}=
\hat{\bbox{p}}_1+\hat{\bbox{p}}_2$,
and relative-motion (rm) coordinates,
$\bbox{r}=\bbox{r}_1-\bbox{r}_2$,
$\hat{\bbox{p}}=\left(\hat{\bbox{p}}_1-\hat{\bbox{p}}_2\right)/2$.
The two-body Hamiltonian thus splits into a CM and a rm part:
\begin{equation}
\hat{\cal{H}}=\hat{H}_{\text{CM}}+\hat{H}_{\text{rm}},
\end{equation}
where
\begin{equation}
\hat{H}_{\text{CM}}=\frac{{\bbox{\hat{P}}}^2}{2M}
+\frac{1}{2}M\omega^2R^2 ,
\label{CMham}
\end{equation}
\begin{equation}
\hat{H}_{\text{rm}}=\frac{{\bbox{\hat{p}}}^2}{2\mu}
+\frac{1}{2}\mu\omega^2r^2+\frac{e^2}{\kappa r},
\label{rmham}
\end{equation}
with $M=2m^{\ast}$, and $\mu=m^{\ast}/2$. The CM hamiltonian
$\hat{H}_{\text{CM}}$ has the form of a simple harmonic oscillator. For the
rm Hamiltonian, $\hat{H}_{\text{rm}}$, it is easy to separate variables and
obtain a radial differential equation, which gives solutions with the same
set of quantum numbers as for the harmonic oscillator. 
Solutions and
notations for the 2D and 3D case are summarized in Appendix A 
for both
$\hat{H}_{\text{CM}}$ and $\hat{H}_{\text{rm}}$.

By denoting the CM and rm quantum numbers with capital and small
letters respectively, the eigenvalues for the two-particle system
can be written as
\begin{equation}
E_{NM,nm}=\hbar\omega\left(2N+\left|M\right|+1\right)+
\epsilon_{nm}
\label{t1}
\end{equation}
in the 2D case, and
\begin{equation}
E_{NL,n\ell}=\hbar\omega\left(2N+L+\frac{3}{2}\right)+
\epsilon_{n\ell}
\label{t2}
\end{equation}
in the 3D spherical case,
the cylindrical 3D Helium QD reducing to an effective 2D one (see
Appendix \ref{appA}).
Here $\epsilon_{nm}$ and $\epsilon_{n\ell}$ are the rm
eigenvalues in 2D and 3D, respectively. Note that degeneracy is strongly
reduced by Coulomb interaction with respect to the non-interacting case.

The corresponding two-particle total eigenfunctions are
\begin{equation}
\Psi_{NM,nm;SS_z}\!\left(\bbox{r}_1,s_1;\bbox{r}_2,s_2\right)=
\Phi_{NM}\!\left(\bbox{R}\right)\varphi_{nm}\!\left(\bbox{r}\right)
\chi\!\left(S,S_z\right);
\label{t3}
\end{equation}
for 2D and the 3D cylinder, and
\begin{equation}
\Psi_{NLM_z,n\ell m_z;SS_z}\!\left(\bbox{r}_1,s_1;\bbox{r}_2,s_2\right)=
\Phi_{NLM_z}\!\left(\bbox{R}\right)\varphi_{n\ell m_z}\!\left(\bbox{r}\right)
\chi\!\left(S,S_z\right);
\label{t4}
\end{equation}
for a 3D sphere.
Here $\Phi\!\left(\bbox{R}\right)$ and $\varphi\!\left(\bbox{r}\right)$ are
respectively the spatial CM and rm
eigenfunctions, and $\chi\!\left(S,S_z\right)$ is the spin function of a
state with total spin $\hbar^2S\left(S+1\right)$ and $z$ projection $S_z$.
Note that the parity of the rm spatial eigenfunction is defined
(total orbital
angular momentum and spin are conserved) and connected with the value
of total spin by the antisymmetry of the two-particle total wavefunction
$\Psi\!\left(\bbox{r}_1,s_1;\bbox{r}_2,s_2\right)$.
For both the disk and the cylinder, this implies
that if $m$ is even, the state is a singlet ($S=0$), and if $m$ is odd, the
state is a triplet ($S=1$). 
Similarly for the sphere case, 
if $\ell$ is even, the state
is a singlet ($S=0$), and if $\ell$ is odd, the state is a triplet ($S=1$).

In the above eigenvalues and eigenfunctions of the two-electron dot, the
ingredients related to the CM Hamiltonian are known analytically (see
Appendix \ref{appA}), while the rm energies
and wavefunctions must be determined
numerically. This is done by exact diagonalization of the rm eigenvalue
problem (Appendix \ref{appA}), thereby yielding the
full 2D and 3D spectrum of the QD helium.

Before  comparing these exact results with the SSH approach,
we point out that the Hamiltonian (\ref{eq:h2e}) can be translated 
into a second-quantized form; this is done in terms of the same
quantum numbers using CM and rm variables. The two-particle Hilbert
space is the Kronecker product of CM and rm single-particle spaces,
generated respectively by the basis $\left\{\left|N\right>\right\}_N$ (with
eigenvalues $E_N$ and creation operators $\hat{a}_N^{\dagger}$) and
$\left\{\left|n\right>\right\}_n$ 
(with eingenvalues $\epsilon_n$ and
creation operators $\hat{a}_n^{\dagger}$). For simplicity, $N$ and $n$
label here the whole set of CM and rm quantum numbers,
respectively. 
The second-quantized form of the two-particle hamiltonian
$\hat{\cal{H}}$, in this variables, is then given by
\begin{equation}
\hat{\cal{H}}=\sum_N E_N \hat{a}_N^{\dagger} \hat{a}_N+
\sum_n \epsilon_n \hat{a}_n^{\dagger} \hat{a}_n+
\sum_{nn^{\prime}}V_{nn^{\prime}}\hat{a}_n^{\dagger}
\hat{a}_{n^{\prime}}.
\label{h:pert}
\end{equation}
This formulation allows us to obtain the result of the previously discussed
Hubbard model, by simply neglecting all off-diagonal matrix elements
in equation (\ref{h:pert}):
\begin{equation}
\hat{\cal{H}}^{SSH} =
\sum_N E_N \hat{a}_N^{\dagger} \hat{a}_N+
\sum_n \left(\epsilon_n+V_{nn}\right) \hat{a}_n^{\dagger} \hat{a}_n,
\label{eq:pert}
\end{equation}
i.e.~manifestly the non interacting Hamiltonian ``renormalized'' by
Coulomb interaction.

In order to check the reliability of the approximations and the role of
dimensionality of the confining potential we have calculated ground state
properties for QD's with different confinement energies, i.e.~different
values of $\hbar \omega_0$, assuming either a 2D or a 3D confining
potential.
The quality of the ground-state eigenfunctions can be probed by the 
spatial pair correlation function
$f\!\left(\bbox{r}\right)$:
\begin{equation}
f\!\left(\bbox{r}\right)= K \left< \sum_{i\neq j}
\delta\left(\bbox{r}-\bbox{r_i+r_j}\right)\right>.
\label{pair}
\end{equation}
Because of the circular symmetry, $f\!\left(\bbox{r}
\right)$ depends only on the modulus of the relative distance $\bbox{r}$.
Here, the factor $K$ is chosen in such a way that, if we define the
dimensionless relative distance $x=r\sqrt{2m^{\ast}\omega_0/\hbar}$,
the quantity $g\!\left(x\right)=xf\!\left(x\right)$ for the 
2D and
the 3D cylinder case,  and $g\!\left(x\right)=x^2 f\!\left(x\right)$
for a 3D sphere is normalized: 
\begin{displaymath}
\int_0^{\infty}\!\!\!g\!\left(x\right){\rm\,d}x=1.
\end{displaymath}
We have calculated this quantity both exactly and according to the SSH scheme
for a in-plane confining energy $\hbar\omega_0$=5 meV
(throughout the paper we use $m^{\ast}=$ 0.065 $m_e$ inside
the dot and $m^{\ast}=$ 0.079 $m_e$ outside, $\kappa=$ 12.98, as
in the QD of ref.~\onlinecite{Tarucha}; $m_e$ is the electronic mass);
the results are shown in Fig.~\ref{fig1}.
The deviations between SSH and exact results clearly depend on the
dimensionality of the confining potential: in disk-shaped 2D QD's the SSH
approximation is found to overestimate the probability of finding 
the two electrons
close together, in analogy with HF results;\cite{Pfannkuche}
the differences between exact and SSH results are
significantly reduced assuming a 3D confining potential.
This result is coherent with what is found for other ground-state
properties: Fig.~\ref{fig2}(a)
 shows the ground-state energies calculated for
dots with different confinement energies, $\hbar \omega_0$, in the
range between 4.5 and 10 
meV. We compare the exact results with the outcomes of HF and SSH
calculations assuming 2D and 3D confinement potentials. 
Notice that the differences between HF and SSH are always smaller 
---by approximately 50\%--- than the corresponding differences 
with respect to the
exact results; moreover the 3D confinement reduces the overall deviation of
both HF and SSH by about 60\%.

Since the SSH scheme is exact at the first perturbative order in the
off-diagonal matrix elements of the e-e interaction,
it is interesting to check the importance of the next perturbative
corrections. Details of how the perturbative expansion is actually
performed for the helium QD are reported in Appendix \ref{perturbation}.
Figure \ref{fig3} reports exact and SSH ground-state energies
compared with the results of second order perturbation theory, showing that
second-order corrections become much smaller if a 3D confinement is assumed.

The situation becomes more complicated when considering dots with smaller
confinement energies: in this case the HF and SSH differ from the exact
result not only quantitatively but also qualitatively, predicting the
two-particle ground state to be a triplet instead of a singlet, as it should
be. This is shown in Fig.~\ref{fig2}(b) where again the exact, HF and SSH
results are shown for dots of different confinement energies.
The difference between triplet and singlet state energies
decreases with increasing confinement energy both for SSH and HF
approximations until a crossover occurs; assuming a 3D confining potential
the confinement energy of this
crossover is reduced, and this again is true
both for SSH and HF approximations.

We may summarize this analysis on helium QD by concluding that the assumed 3D
confinement potential reduces the differences between approximate (SSH and
HF) solutions with respect to the exact ones, both in terms of ground-state
eigenfunctions and eigenenergies.

\section{The many-electron problem}\label{results}

The key quantity that characterizes single-electron
transport into a QD is the addition
energy, i.e., the energy $A\!\left(N\right)$ required to place an extra
electron into a dot that is initially occupied by $N-1$ electrons. Such
quantity, analogous to electron affinity in atomic physics, can be measured
experimentally as a function of $N$. It has been shown\cite{Tarucha} 
that the measured voltage increment $\Delta A$ between
successive single-electron tunneling processes ---i.e., between two
successive maxima in the conductance--- peaks at ``magic'' values of $N$
corresponding to the filling of complete shells ($N = 2, 6, 12$), as well 
as to half-shell filling (e.g. $N = 4$). The existence of these half-shell
filling features is reminiscent of Hund's rule in atomic
physics,\cite{Tarucha,Wang,Wojs} and is intimately related to
electron-electron interaction.

The results of SSH theory for the addition-energy variations, 
$\Delta A\!\left(N\right) = A\!\left(N+1\right)-A\!\left(N\right)$,
are displayed in
Fig.~\ref{fig4} as a function of the electron number $N$ for two different
3D cylindrical quantum dots. Here, $A\!\left(N\right)$ is obtained as
$E^{SSH}\!\left(N\right)-E^{SSH}\!\left(N-1\right)$, where
$E^{SSH}\!\left(N\right)$ is the ground-state energy in Eq.~(\ref{eq:E}). 
As we can see, $\Delta A\!\left(N\right)$ exhibits peaks corresponding both to
complete and half shell filling, thus
well reproducing the experimental evidence in Ref.\onlinecite{Tarucha}.
This behaviour is the result of the interplay between single-particle
contributions and electron-electron repulsion: the single-particle term
favors complete shell filling, while the repulsion among parallel-spin
electrons, smaller than the repulsion among opposite-spin ones, makes the
configurations with maximum total spin energetically favored (Hund's
rule). This is the physical origin of the half shell-filling structure: 
indeed, adding an electron to a half-filled shell forces the double occupancy
of a level; consequently, $\Delta A$ is raised by the dominant Coulomb
repulsion $U_{\alpha\alpha}$ between opposite-spin electrons on the same
level.

For some non-closed shell configurations the total spin turns out to be not
determined by Hund's rule: in particular, for $N$=16 we find a ground state
with total spin $S$=0. Similar deviations
from Hund's rule have been found for large electron numbers
($N>$20) and associated to spin-density-wave instabilities;  for smaller
numbers ($N$=16,18) the same $S$=0
spin-density-wave state has been found to be
a low-energy ``spin isomer'', slightly higher in energy than the ground state
configuration.\cite{Manninen}

From our calculations we may say that the $S$=0 configuration in dots with
large $N$ may be favoured by the reduced repulsion between electrons in high
shells: in the fourth shell, for instance, the Coulomb
integrals $U_{\alpha\beta}$ relative to orbitals with higher values
of the orbital momentum
may be smaller than the corresponding terms relating two levels with smaller
angular momentum;
the double occupation of an orbital with high orbital momentum
$m$ (i.e., the level with $n$=0, $m$=3, see App.~\ref{appA}
for the notation)
with antiparallel-spin
electrons may therefore cost less than having parallel-spin
electrons on different degenerate orbitals, but with
smaller $m$ (i.e., the level $n$=1, $m$=1).
The same interplay also explains the peaks
in $\Delta A\!\left(N\right)$ for $N=$14 and $N$=18.\cite{Notafranca}

We want to stress that also in the case of many electrons the reliability
of the results of SSH approach is comparable with HF ones.
The explicit
comparison between the addition energy variation calculated according to
SSH and HF schemes and for 2D and 3D confinements 
is reported in Fig.~\ref{fig5}, 
showing that $\Delta A$ always peaks at the
same electron numbers and that the agreement between SSH and HF results
improves on going from the 2D to the 3D confinement model.

Ground-state configurations and filling rules change when a magnetic field
is applied. It affects both single particle energies and
Coulomb and exchange integrals through the induced changes in the
wavefunction localization. 
Figure \ref{fig6} shows the $U$- and $J$-
integrals vs.~$B$ for the first states, obtained for
$\hbar\omega_0$=7.5 meV.
For comparison, we also show the corresponding quantities
calculated within a 
strictly 2D-confinement model. 
We can see that $U$-integrals
describing the interaction between opposite-spin electrons are a few meV
smaller in the case of 3D confinement, while the differences in the
interaction between parallel-spin ones are much smaller. This is going to
affect dramatically the energy balance which determines ground-state
configurations, thus clearly showing the failure of a pure 2D description
of state-of-the-art QD structures.

As already mentioned, according to the SSH approach the off-diagonal matrix
elements of the electron-electron interaction are assumed to be negligible.
In Fig.~\ref{fig6} the values of two of them are reported
as functions of the applied magnetic field.
As expected, we clearly see that for
any $B$ value they are negligible 
compared to all the other semi-diagonal contributions, and
even more so in 3D with respect to the 2D case.

Figure \ref{fig7} shows the total energy
$E^{SSH}$ 
as a function of the applied
magnetic field $B$ for different values of the electron number $N$ 
in a dot with confinement
energy $\hbar\omega_0$=7.5 meV. 
It appears that for sufficiently
large values of $B$ the Hamiltonian term 
linear in the magnetic field becomes dominant,
making configurations with higher total angular quantum number
energetically favorable. This is also the physical origin of the wiggles in
the $A\!\left(N\right)$ vs.~$B$ plot shown in
Fig.~\ref{fig8} and observed in the
experiments reported in Ref. \onlinecite{Tarucha}.

Other authors have explicitely considered the question of
dimensionality in theoretical modelization of semiconductor QDs.
Kumar {\em et al.}\cite{Kumar}~computed self-consistently 
the one-particle confining potential in a square QD.
According to their results, our assumption of an in-plane parabolic 
confining potential plus a well in the perpendicular direction is
seen to be quite reliable and general, as well as 
the {\em ansatz} of considering only
the ground state motion along $z$, at least for few electron dots.
Steinebach {\em et al.}\cite{Steinebach}~pointed out the importance 
of a full 3D model to treat spin density excitations (SDEs) in semiconductor
wires and dots. Specifically, they used the analogous of
eq.~(\ref{eq:cdef}) as effective 3D Coulomb interaction, and they found 
that a 2D description artificially enhances the interaction strenght
and is unable to predict experimental Raman spectra.
The necessity of a 3D modelization is then seen to emerge not only in
the description of ground state and single particle processes, like
addition spectra, but also in two particle processes, like SDEs.

\section{Summary and conclusions}\label{conclusions}

We have presented a theoretical investigation of 
Coulomb-correlation effects in semiconductor quantum dots. 
In particular, we have performed a detailed analysis of 
the addition-spectrum problem for 
few-electron quantum-dot structures (macroatoms), 
pointing out possible analogies with more conventional 
Coulomb-correlation effects in atomic physics. 

Our primary goal was to understand to which extent the various 
approximation schemes, such as Hartree-Fock or Hubbard models, are able to 
properly describe Coulomb correlation in realistic, state-of-the-art QDs.
To this end, we have first compared approximate results to the exact 
solution for the prototypical case of a two-electron system, the so-called 
quantum-dot Helium; we have repeated such analysis for different
dimensionalities, considering 
the 3D (spherical and cylindrical geometry) and the pure 2D structures.
The main result is that the degree of accuracy of 
any approximation scheme depends strongly on the dimensionality of 
the problem. More specifically, the pure 2D model ---often used for the 
description of quantum dots--- is found to give 
approximated results which differ significantly from the exact solution. 
We have demonstrated that this 
is not a general failure of the approximation scheme, but it reflects a 
rather pathological behaviour originating from the unphysical nature of the
pure 2D model. Indeed, for the case of a 3D cylindrical model ---which 
provides a much better description of realistic QD structures--- the 
difference between exact solution and approximated results is found to be 
much smaller, thus confirming the validity of the various approximation 
schemes considered.

The same analysis has been then extended to many-electron systems for 
which addition-spectra measurements are available. Using different
approximation schemes, we find that the deviations between the 
full 3D and the simplified 2D quantum-dot model are very significant.
The full 3D model is found to reproduce the experimental data for a 
large class of QD structures where simplified 2D models fail.
We conclude that this is due to the unphysical character of the pure 2D 
confinement for which the various approximation schemes often yield
unreliable results. A proper description 
of the QD structure in terms of fully 3D single-particle wavefunctions
is therefore required; we have shown that 
in this case approximate approaches can give an accurate 
description of correlation effects in the macroatoms made available by 
present semiconductor technology.

We are grateful to D.~Pfannkuche and D.~Vanossi for useful discussions.
This work was supported in part by the EC through the TMR Network 
``Ultrafast quantum optoelectronics''.

\appendix\section{
CM and rm solutions for the two-electron dot}\label{appA}

The present appendix is organized as follows:
In Sec.~\ref{appAcyl} 
we shall show how to reduce the 3D-cylindrical
Helium problem to an effective two-dimensional one;
In Sect.~\ref{appAesatta}
we shall summarize the 2D and 3D solutions
of the one-particle Schr\"odinger equations for the
center-of-mass and the relative-motion Hamiltonians,
$\hat{H}_{\text{CM}}$ and $\hat{H}_{\text{rm}}$,
as defined in eqs.~(\ref{CMham}) and (\ref{rmham}).

\subsection{3D eigenvalue equation for the cylindrical QD}\label{appAcyl}

If only the lowest
single-particle state $\phi_0\!\left(z\right)$ of
the quantum well is relevant
to the two-electron motion, we can write the spatial-part
$\Psi\!\left(\bbox{r}_1;\bbox{r}_2\right)$ of
our Helium wavefunction as
\begin{equation}
\Psi\!\left(\bbox{r}_1;\bbox{r}_2\right)=
\psi\!\left(x_1,y_1;x_2,y_2\right)\phi_0\!\left(z_1\right)
\phi_0\!\left(z_2\right).
\label{eq:Psicyl}
\end{equation}
This approximation is well justified for most cases of interest.
Indeed, for the typical QD structure used in the experimental investigation 
of addition spectra~\cite{Tarucha}
(quantum-well width $L = 12$\,nm, barrier height $V_0 = 200$\,meV, 
the energy separation between ground and first excited state
along $z$ is 56 meV, about one order of magnitude
larger than typical in-plane single particle confinement energies.

Let us now consider the global Schr\"odinger Equation
corresponding to the exact Helium Hamiltonian of Eq.~(\ref{eq:h2e})
\begin{equation}
\hat{\cal{H}}\Psi\!\left(\bbox{r}_1;\bbox{r}_2\right)=
E\Psi\!\left(\bbox{r}_1;\bbox{r}_2\right);
\label{eq:Scglobal}
\end{equation}
by substituting  eq.~(\ref{eq:Psicyl}), multiplying both sides
by $\phi_0^{\ast}\!\left(z_1\right)
\phi_0^{\ast}\!\left(z_2\right)$ and integrating over $z_1$
and $z_2$, we obtain:
\begin{equation}
\left[2\varepsilon^z_0+\sum_{i=1}^2\hat{H}_0\left(i\right)+
\frac{e^2}{\kappa}c\!\left(\left|\bbox{r_1-r_2}\right|\right)\right]
\psi\!\left(\bbox{r}_1;\bbox{r}_2\right)=
E\psi\!\left(\bbox{r}_1;\bbox{r}_2\right).
\label{eq:Auglobal}
\end{equation}
The eigenvalue equation is then reduced to a 2D one,
since $\bbox{r}_i\equiv\left(x_i,y_i\right)$ and
$c\!\left(r\right)=
c\!\left(\left|\bbox{r_1-r_2}\right|\right)$ is an
{\em effective} Coulomb potential, accounting for 
the geometry of the system:
\begin{equation}
c\!\left( r\right)=
\int_{-\infty}^{+\infty}\!\!\!{\rm\,d}z_1
\int_{-\infty}^{+\infty}\!\!\!{\rm\,d}z_2
\frac{\left|\phi_0^{\ast}\!\left(z_1\right)\right|^2
\left|\phi_0^{\ast}\!\left(z_2\right)\right|^2}
{\sqrt{r^2 +
\left(z_1-z_2\right)^2}};
\label{eq:cdef}
\end{equation}
From now on we will drop the constant ground-state energy
along $z$ ($\varepsilon^z_0$).
As a first step, we evaluate $c\!\left(r\right)$ 
by solving the quantum well eigenvalue problem 
(allowing for different values of the effective mass 
in the well and in the barrier); 
Then, we numerically integrate eq.~(\ref{eq:cdef}).
It is easy to show analytically some important properties
of $c\!\left(r\right)$, namely that
\begin{equation}
0 \leq r c\!\left(r\right) \leq 1 \qquad \forall r,
\end{equation}
\begin{equation}
\lim_{r\to 0}\,  r c\!\left(r\right) = 0,
\end{equation}
\begin{equation}
\lim_{r\to \infty}\,  r c\!\left(r\right) = 1.
\end{equation}
These properties tell us that for large distances $r$
$c\!\left(r\right)$ tends to the bare Coulomb potential, and that
it is however strongly reduced 
in the neighborhood of the origin, 
i.e., the more relevant space region in the computation of 
Coulomb and exchange integrals.
Figure \ref{fig9} shows such effective Coulomb potential $C$ 
multiplied by 
the dimensionless variable $x$ (introduced below) as a function of $x$  
for different values of the quantum-well width: 
A monotonous behavior is apparent, going from
the bare Coulomb-potential value in the zero-width limit
(the function is constant and equal to 1), into progressively smaller values,
towards the infinite-width case.

\subsection{Exact solutions}\label{appAesatta}

Let us first consider the CM equation, which
has the form of a standard harmonic oscillator and can thus
be solved analytically.
For the 2D case (3D cylindrical), its eigenvalues are:
\begin{eqnarray}
\varepsilon^{\text{2D}}_{NM}&=&
\hbar\omega_0\left(2N+\left|M\right|+1\right)\nonumber\\
N=0,1,2,\ldots&&M=0,\pm1,\pm2,\ldots
\label{fo:2D}
\end{eqnarray}
and the corresponding ortonormalized eigenfunctions
(the so called ``Fock-Darwin'' states\cite{rev_th,Fock}) are
\begin{equation}
\Phi^{\text{2D}}_{NM\sigma}\!\left(\bbox{r},s\right)=
\left<s,\bbox{r}|NM\sigma\right>=
\lambda^{\frac{\left|M\right|+1}{2}}\sqrt{\frac{N!}
{\pi\left(N+\left|M\right|\right)!}}\,{\rm e}^{-{\rm i}M\varphi}
r^{\left|M\right|}{\rm e}^{-\frac{\lambda r^2}{2}}
L_N^{\left|M\right|}\!
\left(\lambda r^2\right)\chi_{\sigma}\!\left(s\right)
.
\label{eigen2D}
\end{equation}
In the 3D spherical case,\cite{Flugge} the eigenvalues are
\begin{eqnarray}
\varepsilon^{\text{3D}}_{NL}&=&
\hbar\omega_0\left(2N+L+\frac{3}{2}\right)\nonumber\\
N=0,1,2,\ldots&&L=0,1,2,\ldots
\label{fo:3D}
\end{eqnarray}
and the ortonormalized eigenfunctions
$\Phi^{\text{3D}}_{NL M_z\sigma}\!\left(\bbox{r},s\right)$ are
\begin{equation}
\Phi^{\text{3D}}_{NM L_z\sigma}\!\left(\bbox{r},s\right)=
\left<s,\bbox{r}|NL M_z\sigma\right>=
\sqrt{\frac{2\lambda^{L+\frac{3}{2}}N!}
{\Gamma\!\left(N+\L+\frac{3}{2}\right)}}\,
r^{L}{\rm e}^{-\frac{\lambda r^2}{2}}
L_N^{L+\frac{1}{2}}\!\left(\lambda r^2\right)
Y_{L M_z}\!\left(\vartheta,\varphi\right)
\chi_{\sigma}\!\left(s\right).
\label{eigen3D}
\end{equation}
Here, $\lambda = \frac{m^{\ast}\omega_0}{\hbar}$,
$L_N^p$ are generalized Laguerre polynomials,\cite{Abramowitz} 
$\Gamma$ is the usual Gamma function,
$\chi_{\sigma}$ denotes the spin function,
and $Y_{L M_z}$ are the spherical harmonics.
We have used polar coordinates throughout:
$\bbox{r}\equiv\left(r,\varphi\right)$ in 2D (3D cylindrical),
case and $\bbox{r}\equiv\left(r,\vartheta,\varphi\right)$ in the 3D spherical 
case.
For the 2D (3D cylindrical)
the quantum numbers are $\left(N,M,\sigma\right)$:
$N$ is the radial quantum number, $M$ the angular momentum
quantum number (in this case the total angular momentum
coincides with the component along $z$, $L_z=-\hbar M$), and
$\sigma$ the spin component along $z$.
In the 3D spherical case, on the other hand, the quantum numbers are given 
by
$\left(N,L,M_z,\sigma\right)$: 
here $L$ is the total
angular momentum quantum
number, and $M_z$ is the magnetic quantum number, $M_z=-L,-L+1,
\ldots,L$.

Let us now come to the single-particle Schr\"odinger equation for
the rm Hamiltonian of Eq.~(\ref{rmham}). 
In this equation
the variables are easily separable, and the problem is reduced to
the solution of a radial differential equation.
For the 2D (3D cylindrical) case, the rm eigenfunction in coordinate space is
\begin{equation}
\varphi_{nm}\!\left(\bbox{r}\right)=R_{nm}\!\left(r\right)
\frac{{\rm e}^{-{\rm i}m\varphi}}{\sqrt{2\pi}},
\end{equation}
where 
$R_{nm}\!\left(r\right)$
is the solution of the radial Schr\"odinger equation
\begin{equation}
\frac{{\partial}^2R_{nm}
\!\left(r\right)}{\partial r^2}+\frac{1}{r}
\frac{\partial R_{nm}\!\left(r\right)}{\partial r}+
\left[k^2_{nm}-{\tilde{\lambda}}^2
r^2-\alpha c\!\left(r\right)-\frac{m^2}{r^2}\right]
R_{nm}\!\left(r\right)=0;
\label{eq2D}
\end{equation}
we have employed the notations $\tilde{\lambda}=\mu\omega_0/\hbar$,
$\alpha=2\mu e^2 / \kappa{\hbar}^2$,
and $k^2_{nm}=2\mu\epsilon_{nm}/{\hbar}^2$, where $\epsilon_{nm}$
is the rm eigenvalue. The effective Coulomb potential
$c\!\left(r\right)$ is simply $1/r$ in the 2D case and it is defined
in Sect.~ \ref{appAcyl} for the 3D cylindrical case.

For the 3D spherical case, the rm eigenfunction in coordinate space is
\begin{equation}
\varphi_{n\ell m_z}\!\left(\bbox{r}\right)=R_{n\ell}\!\left(r\right)
Y_{\ell m_z}\!\left(\vartheta,\varphi\right),
\end{equation}
with $R_{n\ell}\!\left(r\right)$ satisfying the radial eigenvalue
equation
\begin{equation}
\frac{{\partial}^2R_{n\ell}\!\left(r\right)}{\partial r^2}+\frac{2}{r}
\frac{\partial R_{n\ell}\!\left(r\right)}{\partial r}+
\left[k^2_{n\ell}-{\tilde{\lambda}}^2
r^2-\frac{\alpha}{r}-\frac{\ell\left(\ell+1\right)}{r^2}\right]
R_{n\ell}\!\left(r\right)=0;
\label{eq3D}
\end{equation}
where, again, we put
$k^2_{n\ell}=2\mu\epsilon_{n\ell}/{\hbar}^2$, and $\epsilon_{n\ell}$
is the rm eigenvalue.

In order to obtain an exact solution for the rm eigenvalue problems,
we rewrite Eqs.~(\ref{eq2D})
and (\ref{eq3D}) in
terms of the dimensionless variable $x = {\tilde{\lambda}}^{1/2}r$.
For the 2D (3D cylindrical) case, Eq.~(\ref{eq2D})
becomes
\begin{displaymath}
\frac{{\rm d}}{{\rm d}x}\left(x
\frac{{\rm d}{\tilde{R}}_{nm}\!\left(x\right)}{{\rm d}x}\right)
+\left[-\frac{m^2}{x}-\tilde{\alpha}\,x\,C\!\left(x\right)
+{\tilde{k}}^2_{nm}x-x^3
\right]{\tilde{R}}_{nm}\!\left(x\right)=0,
\end{displaymath}
\begin{displaymath}
{\tilde{R}}_{nm}\!\left(x\right)=R_{nm}\!\left(r\right),
\end{displaymath}
\begin{equation}
C\!\left(x\right)={\tilde{\lambda}}^{-1/2}\,
c\!\left({\tilde{\lambda}}^{-1/2}x\right)
\label{eqa2D}
\end{equation}
(for the 2D case it is simply $C\!\left(x\right)=1/x$),
while for the 3D case Eq.~(\ref{eq3D}) transforms into
\begin{displaymath}
\frac{{\rm d}^2
\tilde{\chi}_{n\ell}\!\left(x\right)}{{\rm d}x^2}
+\left[-\frac{\ell\left(\ell+1\right)}{x^2}
-\frac{\tilde{\alpha}}{x}+{\tilde{k}}^2_{n\ell}-x^2
\right]{\tilde{\chi}}_{n\ell}\!\left(x\right)=0,
\end{displaymath}
\begin{equation}
{\tilde{\chi}}_{n\ell}\!\left(x\right)=\chi_{n\ell}\!\left(r\right)
\qquad \chi_{n\ell}\!\left(r\right)=\frac{
R_{n\ell}\!\left(r\right)}{r}.
\label{eqa3D}
\end{equation}
The dimensionless parameters are $\tilde{\alpha}
={\tilde{\lambda}}^{-1/2}\alpha=2\sqrt{{\cal R}^{\ast}/\hbar\omega_0}$,
$\tilde{k}^2_{\alpha}=k^2_{\alpha}/\tilde{\lambda}=
2\epsilon_{\alpha}/\hbar\omega$;
${\cal R}^{\ast}=e^4m^{\ast}/2{\kappa}^2{\hbar}^2$ is the effective
Rydberg energy.
Actually, exact analytic solutions exist,
but they are limited to 2D and 3D
spherical cases only;\cite{Taut,Dineykhan} 
thus, we have chosen to solve Eqs.~(\ref{eqa2D}) and (\ref{eqa3D})
by standard numerical methods.
We stress that the numerical accuracy depends
on the accurate specification of the boundary conditions, that
we impose through analytical asymptotic formulae 
for eigenfunctions near to
the singular points $0$ and $+\infty$, following
the general methods of Ref.~\onlinecite{Moon}.
In this way the numerical solution is very stable and efficient, thus
overcoming possible difficulties related to the singlet
ground state;\cite{Pfannkuche}
in our calculations energy values are obtained with a nominal relative error of
the order of $10^{-8}$.

\section{Helium Perturbation Theory}\label{perturbation}

We employ the standard Rayleigh-Schr\"odinger perturbation theory
to correct the SSH eigenvalues in Eq.~(\ref{eq:pert})
at the second order in the off-diagonal Coulomb matrix elements
entering the total Hamiltonian (\ref{h:pert}).
In the remaining part of this section we shall consider
the 2D and 3D spherical cases, by neglecting the center-of-mass motion.

For the 2D case, the rm SSH eigenvalues
$\epsilon_{nm}^{SSH}$ are given by~\cite{Merkt}
\begin{equation}
\epsilon_{nm}^{SSH}=
\hbar\omega_0\left(2n+\left|m\right|+1\right)+
\sqrt{ {\cal{R}}^{\ast} \hbar\omega_0}
\,\,S^{\text{2D}}\!\left(n,m\right),
\end{equation}
\begin{equation}
S^{\text{2D}}\!\left(n,m\right)=\frac{\Gamma\!\left(\left|m\right|+
\frac{1}{2}\right)}{\left|m\right|!}\left\{
1+\sum_{s=0}^{n-1}\frac{n!
\left(-1\right)^{s+1}\left[\left(2s+1\right)!!\right]^2\left|
m\right|!}
{\left(n-s-1\right)!\,
2^{2s+2}\left[\left(s+1\right)!\right]^2\left(\left|m\right|+s+1
\right)!}\right\};
\label{S2D}
\end{equation}
while for the 3D case we have
\begin{equation}
\epsilon_{n\ell}^{SSH}=\hbar\omega_0\left(2n+\ell+\frac{3}{2}\right)
+\sqrt{ {\cal{R}^{\ast}} \hbar\omega_0}\,\,
S^{\text{3D}}\!\left(n,\ell\right),
\end{equation}
\begin{equation}
S^{\text{3D}}\!\left(n,\ell\right)=\frac{\ell!\,\Gamma\!\left(
\frac{1}{2}\right)\left(2\ell+1\right)!!}{2^{\ell+1}\Gamma^2\!\left(
\ell+\frac{3}{2}\right)}\left\{1+\sum_{s=0}^{n-1}
\frac{n!
\left(-1\right)^{s+1}\Gamma\!\left(\frac{1}{2}\right)\left[
\left(2s+1\right)!!\right]^2\left(2\ell+1\right)!!}
{2^{s+\ell+3}\left(n-s-1\right)!\,
\Gamma\!\left(s+\ell+\frac{5}{2}\right)
\left[\left(s+1\right)!\right]^2}\right\}.
\label{S3D}
\end{equation}

The first-order correction due to non-diagonal Coulomb matrix elements
is equal to zero. 
The second-order correction $\Delta\varepsilon_0^{\left( 2 \right)}$
to the ground-state energy is given
by the well-known expression
\begin{equation}
\Delta\varepsilon_0^{\left(2\right)}=
\sum_n\frac{ \left|V_{0n}\right|^2}{\varepsilon_0^{SSH}-
\varepsilon_n^{SSH}}
\label{eq:IIpert}
\end{equation}
(see notation in eq.~(\ref{h:pert})). 
The idea now is to look for
analytic expressions for the off-diagonal integrals $V_{0n}$ and
then to perform a numerical summation. 
However, expressions like those
obtained in Eqs.~(\ref{S2D})-(\ref{S3D})~\cite{Hawrylacco}
are not useful, since each integral is given by an
alternated-sign summation and numerical errors become rapidly 
critical as the quantum number $n$ increases. 
In contrast, the solution can be obtained using
an integration trick suggested in Ref.~\onlinecite{Stone}, 
so that all the terms in the summation are obtained with the same sign. 
For the 2D case one gets
\begin{displaymath}
V_{n0}=\sqrt{ {\cal{R}^{\ast}} \hbar\omega_0}\,\,
\frac{\Gamma\!\left(n+\frac{1}{2}\right)}{\Gamma\!
\left(n+1\right)},
\end{displaymath}
\begin{equation}
V_{n0}\approx \sqrt{ {\cal{R}}^{\ast} \hbar\omega_0}\,\,
\frac{1}{n^{\frac{1}{2}}} \qquad n \to \infty.
\label{eq:vn02d}
\end{equation}
For the 3D case one obtains
\begin{displaymath}
V_{n0}=\sqrt{ {\cal{R}}^{\ast} \hbar\omega_0}\,\,
\sqrt{\frac{\Gamma^2\!\left(n+\frac{1}{2}\right)}{\pi\,
\Gamma\!\left(n+1\right)\,\Gamma\!\left(n+\frac{3}{2}\right)\,
\Gamma\!\left(\frac{3}{2}\right)}},
\end{displaymath}
\begin{equation}
V_{n0}\approx \sqrt{ {\cal{R}^{\ast}} \hbar\omega_0}\,\,
\frac{1}{\sqrt{\pi\,\Gamma\!\left(\frac{3}{2}\right)}}\,\,
\frac{1}{n^{\frac{3}{4}}}\qquad n\to\infty.
\label{eq:vn03d}
\end{equation}
As already pointed out, 
now the generic terms (\ref{eq:vn02d})-(\ref{eq:vn03d}) in
the sum (\ref{eq:IIpert}) have the same sign and the summation 
can be easily performed. The result for the 2D case is
\begin{equation}
\Delta\varepsilon_0^{\left(2\right)}=
-{\cal{R}}^{\ast}\,\,(0.691),
\end{equation}
and for the 3D case
\begin{equation}
\Delta\varepsilon_0^{\left(2\right)}=
-{\cal{R}}^{\ast}\,\,(0.156).
\end{equation}
Note that the 3D term is significantly smaller than the
corresponding 2D one, and that in the 3D case the series converges faster.

\clearpage
\clearpage
%
%
\begin{figure}
\caption{
Ground-state spatial pair-correlation function $g\!\left(x\right)$ for the
3D (spherical) and 2D two-electron QD: exact and SSH results are reported.
Here, 
$x= \left({2m^{\ast}\omega_0 r^2\over \hbar}\right)^{1 \over 2}$
is the dimensionless relative radial coordinate and
$g\!\left(x\right)$ is normalized in such a way that
$\int_0^{\infty}\!\!\!g\!\left(x\right){\rm\,d} x = 1$. 
The in-plane confinement
energy is $\hbar\omega_0 = 5$\,meV.
}
\label{fig1}
\end{figure}
\begin{figure}
\caption{
(a) Ground-state energy of the artificial He QD as a function of the 
confinement energy $\hbar\omega_0$,
calculated within different approaches (exact, SSH, and HF).
The range of  $\hbar\omega_0$ is 4.5$\div$10 meV.
The two panels correspond to 2D and 3D (cylindrical) geometries.
The ground-state configuration is always a spin-singlet.
(b) Spin-singlet (spin-triplet) energies 
versus confinement energies $\hbar\omega_0$.
The range of  $\hbar\omega_0$ is 1$\div$4.5 meV.
The two panels are relative to 2D and 3D (cylindrical) geometries.
The exact ground state is always a singlet, while a singlet-triplet
crossover occurs for both approximated schemes in the low-energy region.
}
\label{fig2}
\end{figure}
\begin{figure}
\caption{
Ground-state energy of the artificial He QD as a function of the 
confinement energy $\hbar\omega_0$,
as obtained via exact diagonalization, SSH approximation scheme, 
and Rayleigh-Schr\"odinger perturbation theory at the second order
in the off-diagonal Coulomb matrix elements.
Both 3D (spherical) and 2D cases are shown.
}
\label{fig3}
\end{figure}
\begin{figure}
\caption{
Calculated SSH addition-energy increment
$\Delta A$ as a function of the total number $N$ of electrons
for two different QD
structures, both characterized by a parabolic potential
in the $xy$ plane (confining energy $\hbar\omega_0$) and by a
finite-barrier
quantum-well potential along the $z$ direction (3D cylindrical model).
}
\label{fig4}
\end{figure}
\begin{figure}
\caption{
Comparison between SSH and HF addition-energy increments
$\Delta A$ as a function of the total number $N$ of electrons in the dot.
Here, the upper panel corresponds to the 2D geometry while the 
lower one 
corresponds to
the 3D cylindrical model.
The in-plane confinement energy is $\hbar\omega_0 = 7.5$\,meV.
}
\label{fig5}
\end{figure}
\begin{figure}
\caption{
Coulomb ($U_{\alpha;\beta}$) and exchange ($J_{\alpha;\beta}$) integrals
as well as off-diagonal Coulomb matrix elements
($V_{\alpha,\beta;\gamma,\delta}$), as functions of the magnetic field $B$
for both 2D and 3D cases.
Here, $\alpha$ and $\beta$ denote the sets of
radial and angular quantum numbers ($n,m$) for the various single-particle
states
involved in the two-body interaction process.
The in-plane confinement energy is $\hbar\omega_0 = 7.5$\,meV.
}
\label{fig6}
\end{figure}
\begin{figure}
\caption{
Total energy $E^{SSH}$
as a function of the applied magnetic field $B$ 
corresponding to $N$ electrons in a dot with confinement energy
$\hbar\omega_0 = 7.5$\,meV.
For any given value of $N$, all the possible configurations,
denoted by the usual atomic physics terms ${}^{2S+1}L$,
have been considered.
}
\label{fig7}
\end{figure}
\begin{figure}
\caption{
Addition energy $A\!\left(N\right)$
as a function of the magnetic field $B$ 
calculated for a
realistic (3D) QD structure with confinement energy
$\hbar\omega_0 = 7.5$\,meV and for different values of $N$.
The labels indicate the electronic terms for the ground-state
configurations, that depend on $B$.
}
\label{fig8}
\end{figure}
\begin{figure}
\caption{
Plot of the effective Coulomb potential $C\!\left(x\right)$ 
multiplied by the dimensionless coordinate $x$  
($= x\,C\!\left(x\right)$) as a function of $x$  
for different values of the quantum well width $L$
and for a confinement energy $\hbar\omega_0=5$\,meV.
Notice that in the limit
$L \to 0$ (2D case)
$C\!\left(x\right) \to 1/x$ and, therefore,  
$x\,C\!\left(x\right) \to 1$.
}
\label{fig9}
\end{figure}


\begin{references}
\bibitem{rev_exp} For recent reviews see 
R.C.~Ashoori, Nature {\bf 379}, 413 (1996);
L.P.~Kouwenhoven et al., in {\it 
Mesoscopic Electron Transport}, edited by L.L.~Sohn et al. (Kluwer, 
Dordrecht, 1997), p. 105.
\bibitem{Tarucha}
S.~Tarucha, D.G.~Austing, T.~Honda, R.J.~van der Hage,
and L.P.~Kouwenhoven,
Phys.~Rev.~Lett.~{\bf 77}, 3613 (1996).
\bibitem{Fricke}
M.~Fricke, A.~Lorke, J.P.~Kotthaus, G.~Medeiros-Ribeiro,
and P.M.~Petroff,
Europhys.~Lett.~{\bf 36}, 197 (1996).
\bibitem{rev_th} For a recent review see
L.~Jacak, P.~Hawrylak, A.~W\'ojs, 
{\it Quantum Dots} (Springer, Berlin, 1998); 
see also N.F.~Johnson and M.~Reina,
J.~Phys.: Condensed Matter {\bf 4}, L623 (1992);
J.J.~Palacios et al., Phys.~Rev.~B {\bf 50}, 5760 (1994);
and references therein.
\bibitem{noi}
M.~Rontani, F.~Rossi, F.~Manghi, and E.~Molinari,
Appl.~Phys.~Lett.~{\bf 72}, 957 (1998).
\bibitem{threedim} 
A few examples of 3D calculations exist in the literature;
see e.g. Y.~Tanaka and H.~Akera, Phys.~Rev.~B {\bf 53}, 3901 (1996);
A.~Wojs et al., Phys.~Rev.~B {\bf 54}, 5604 (1996);
T.~Ezaki et al., Phys.~Rev.~B {\bf 56}, 6428 (1997).
\bibitem{Pfannkuche}
D.~Pfannkuche, V.~Gudmundsson, and P.A.~Maksym,
Phys.~Rev.~B {\bf 47}, 2244 (1993).
\bibitem{Hubbard}
J.~Hubbard, Proc.~Roy.~Soc.~{\bf A276}, 238 (1963).
\bibitem{Roothaan}
These are the so-called Roothaan equations:
see e.g.~I.~Levine, {\em Quantum Chemistry} (Prentice-Hall,
New Jersey, 1991).
For applications to 2D quantum dots see Refs. \onlinecite{Pfannkuche},
\onlinecite{Wang}.
\bibitem{Merkt}
U.~Merkt, J.~Huser, and M.~Wagner, Phys.~Rev.~B {\bf 43}, 7320 (1991).
\bibitem{Wang}
L.~Wang, J.K.~Zhang, and A.R.~Bishop, Phys.~Rev.~Lett.~{\bf 73},
585 (1994).
\bibitem{Wojs}
A.~W\'ojs and P.~Hawrylak, Phys.~Rev.~B {\bf 53}, 10841 (1996).
\bibitem{Manninen}
M.~Koskinen, M.~Manninen, S.M.~Reimann, Phys.~Rev.~Lett.~{\bf 79},
1389 (1997).
\bibitem{Notafranca}
Since in these cases the energy balance defining the ground state
energy according to Eq.~(\ref{eq:E}) becomes very delicate, the dependence
of Coulomb and exchange integrals on the dot dimensionality may give rise
to different ground state spin polarization in 2D and 3D QD's.
\bibitem{Kumar}
A.~Kumar, S.E.~Laux, and F.~Stern, Phys.~Rev.~B {\bf 42,} 5166 (1990).
\bibitem{Steinebach}
C.~Steinebach, C.~Sch\"uller, G.~Biese, D.~Heitmann, and K.~Eberl,
Phys.~Rev.~B {\bf 57,} 1703 (1998).
\bibitem{Fock}
V.~Fock, Z.~Phys.~{\bf 47}, 446 (1928).
\bibitem{Flugge}
S.~Fl\"ugge, {\em Practical Quantum Mechanics} (Springer-Verlag,
Berlin, 1971).
\bibitem{Abramowitz}
{\em Handbook of Mathematical Functions}, edited by
M.~Abramowitz and I.~A.~Stegun (Dover, New York, 1972).
\bibitem{Taut}
M.~Taut, Phys.~Rev.~B {\bf 48}, 3561 (1993).
\bibitem{Dineykhan}
M.~Dineykhan and R.G.~Nazmitdinov, Phys.~Rev.~B {\bf 55,} 13707 (1997).
This paper also consider different in-plane and orthogonal single particle
parabolic confinement potentials.
\bibitem{Moon}
P.~Moon and D.E.~Spencer, {\em Field Theory Handbook}
(Springer-Verlag, Berlin, 1961).
\bibitem{Hawrylacco}
A.~W\'ojs and P.~Hawrylak, Phys.~Rev.~B {\bf 51}, 10880 (1995);
E.~Anisimovas and A.~Matulis, J.~Phys.~Cond.~Mat.~{\bf 10}, 601 (1998).
\bibitem{Stone}
M.~Stone, H.W.~Wyld, and R.L.~Schult, Phys.~Rev.~B {\bf 45},
14156 (1992).
\end{references}
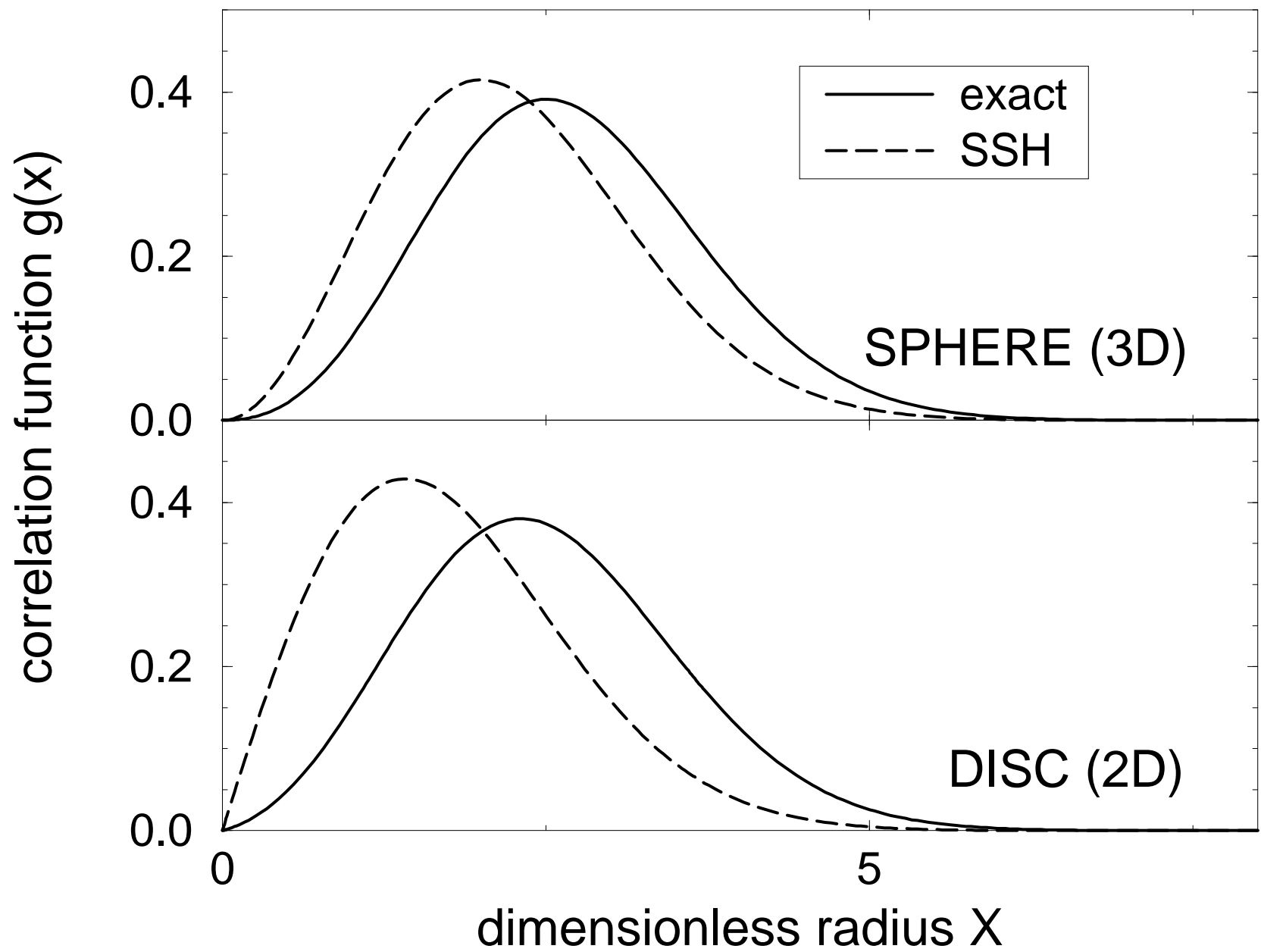
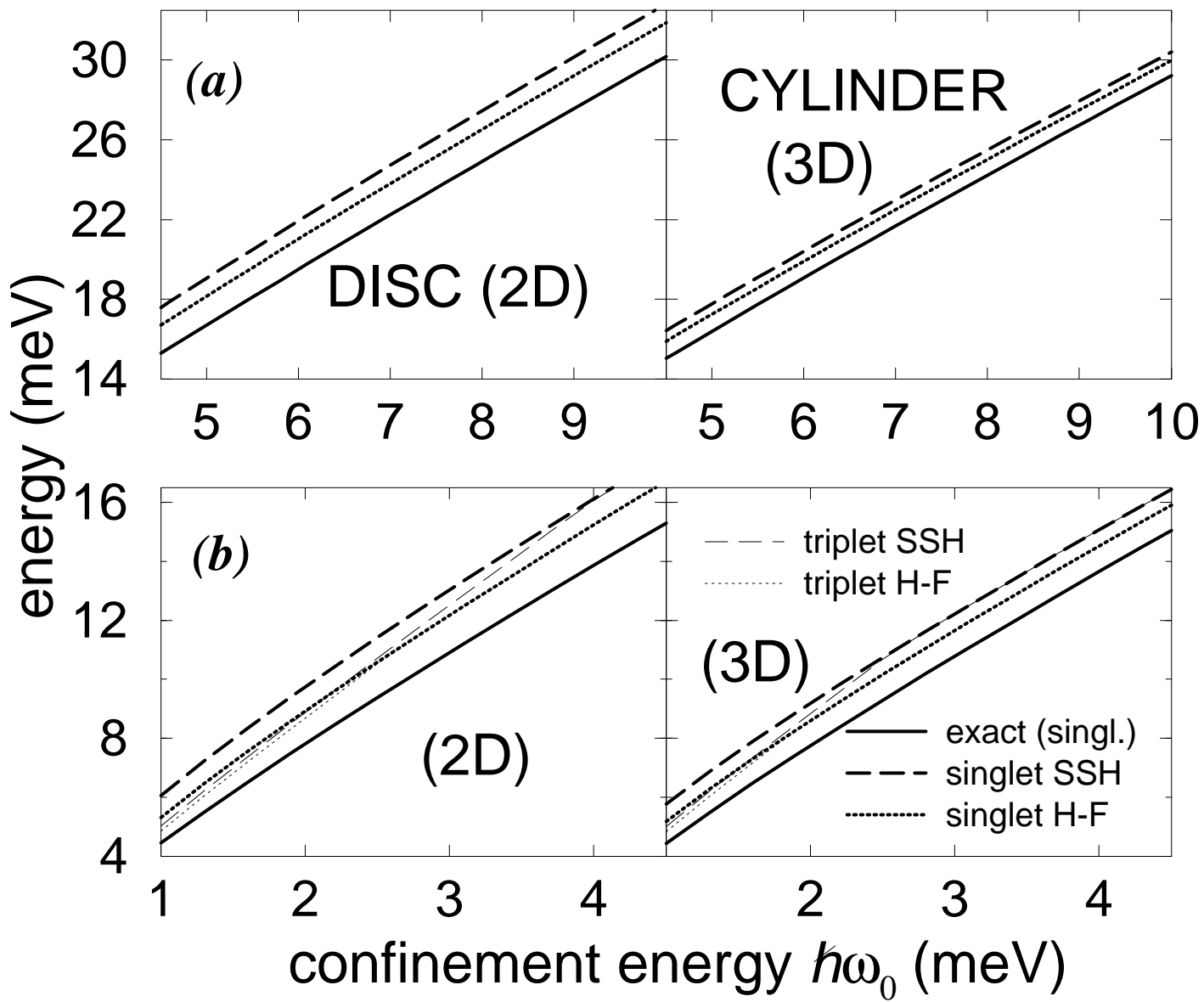
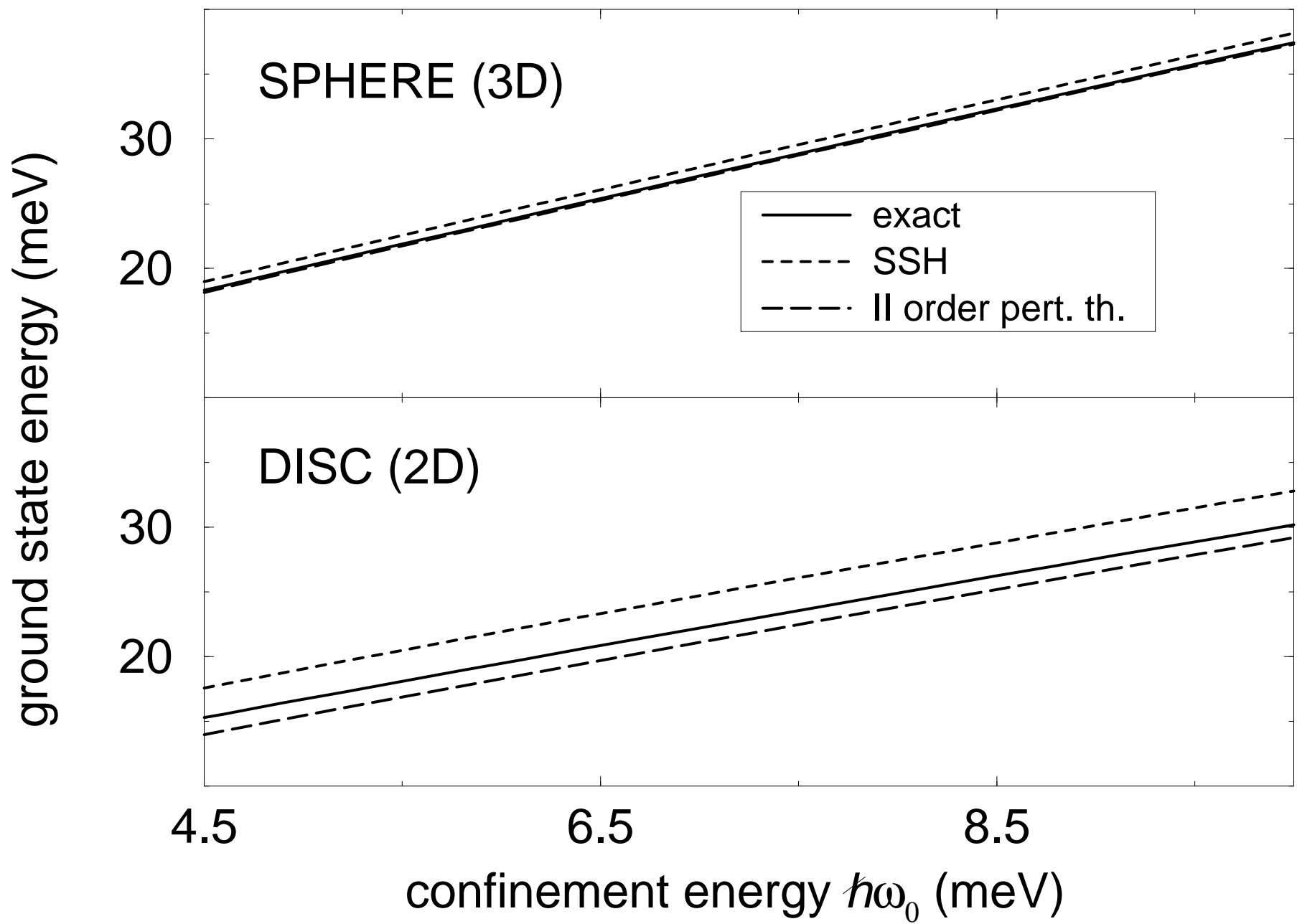
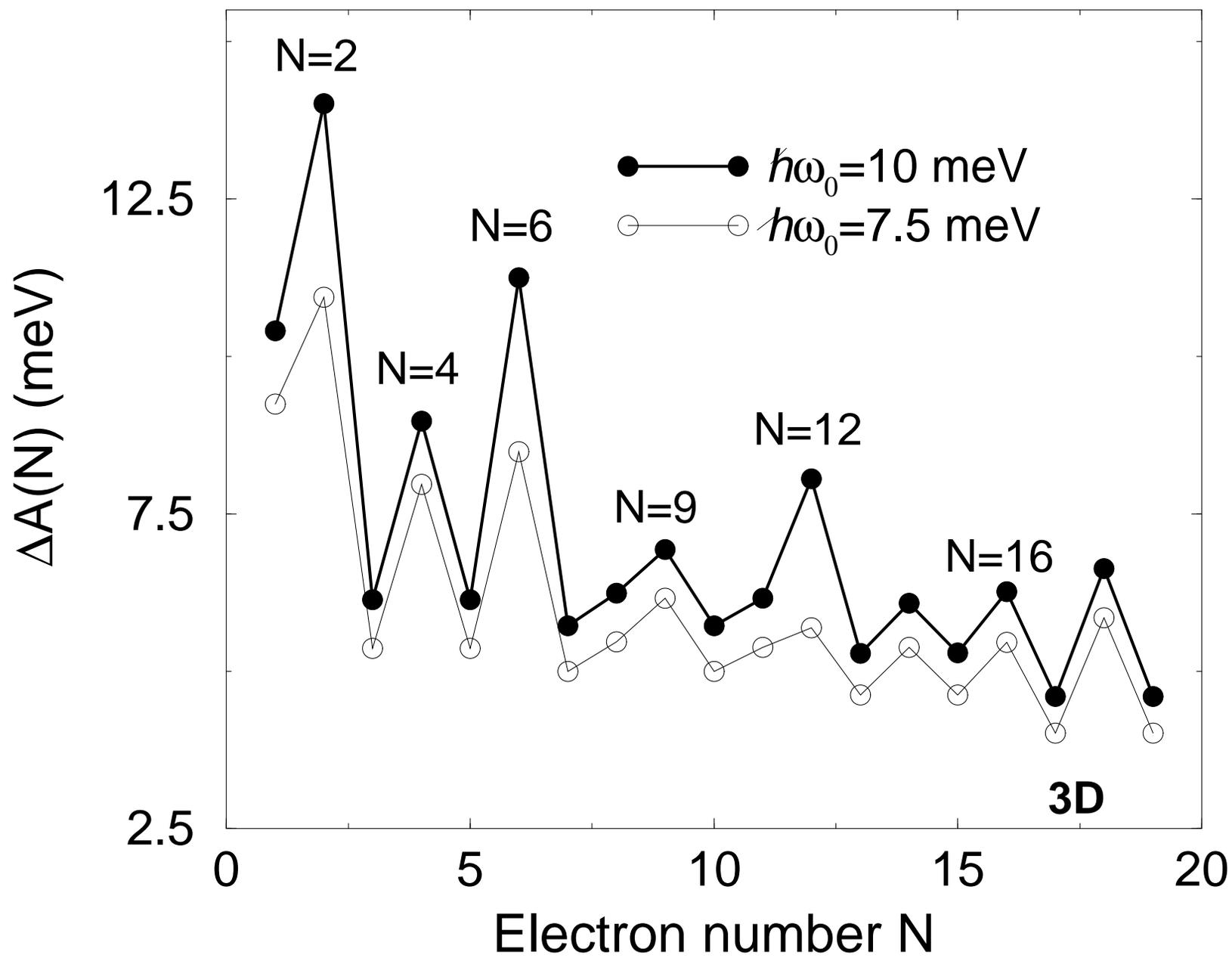
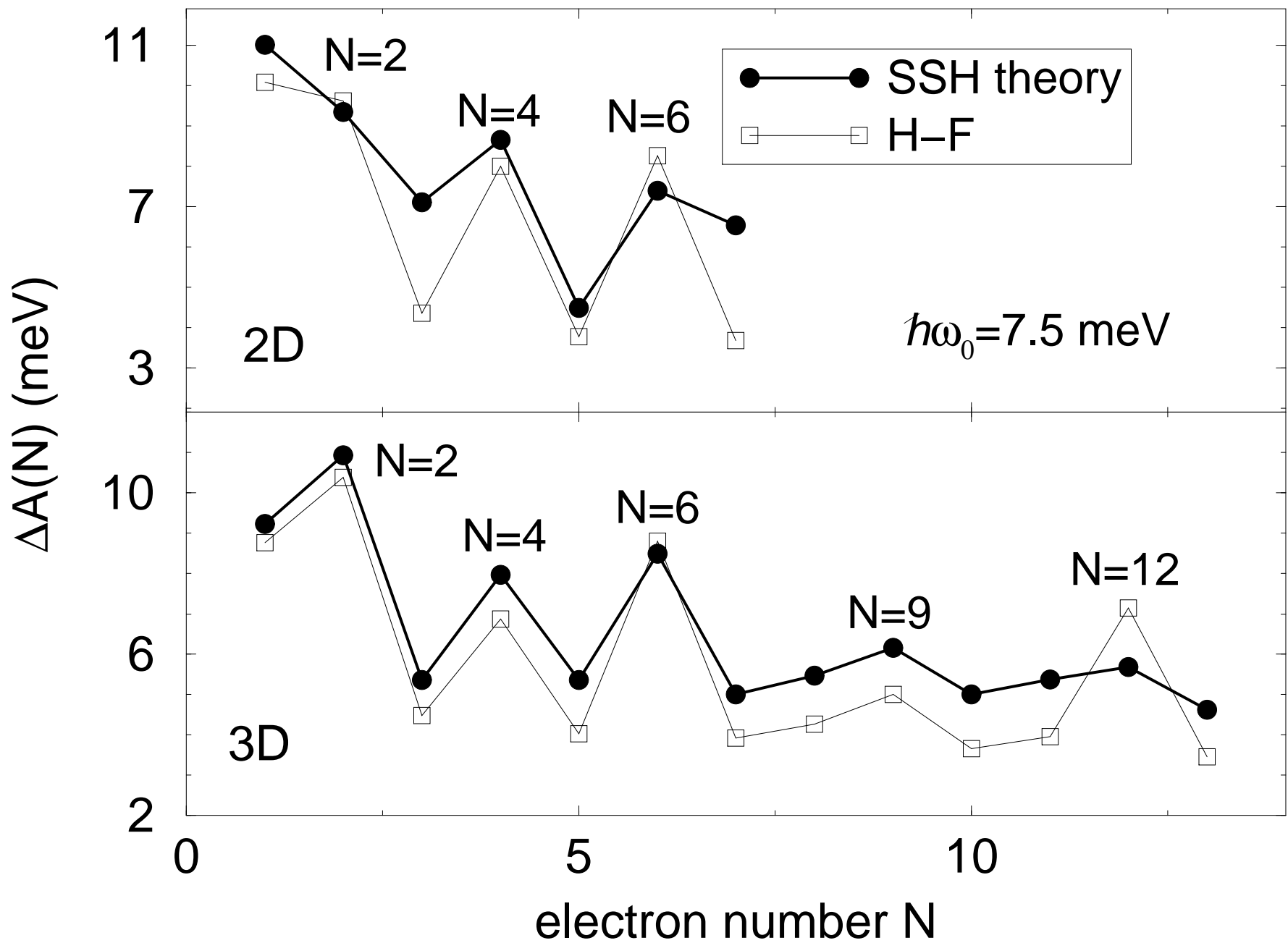
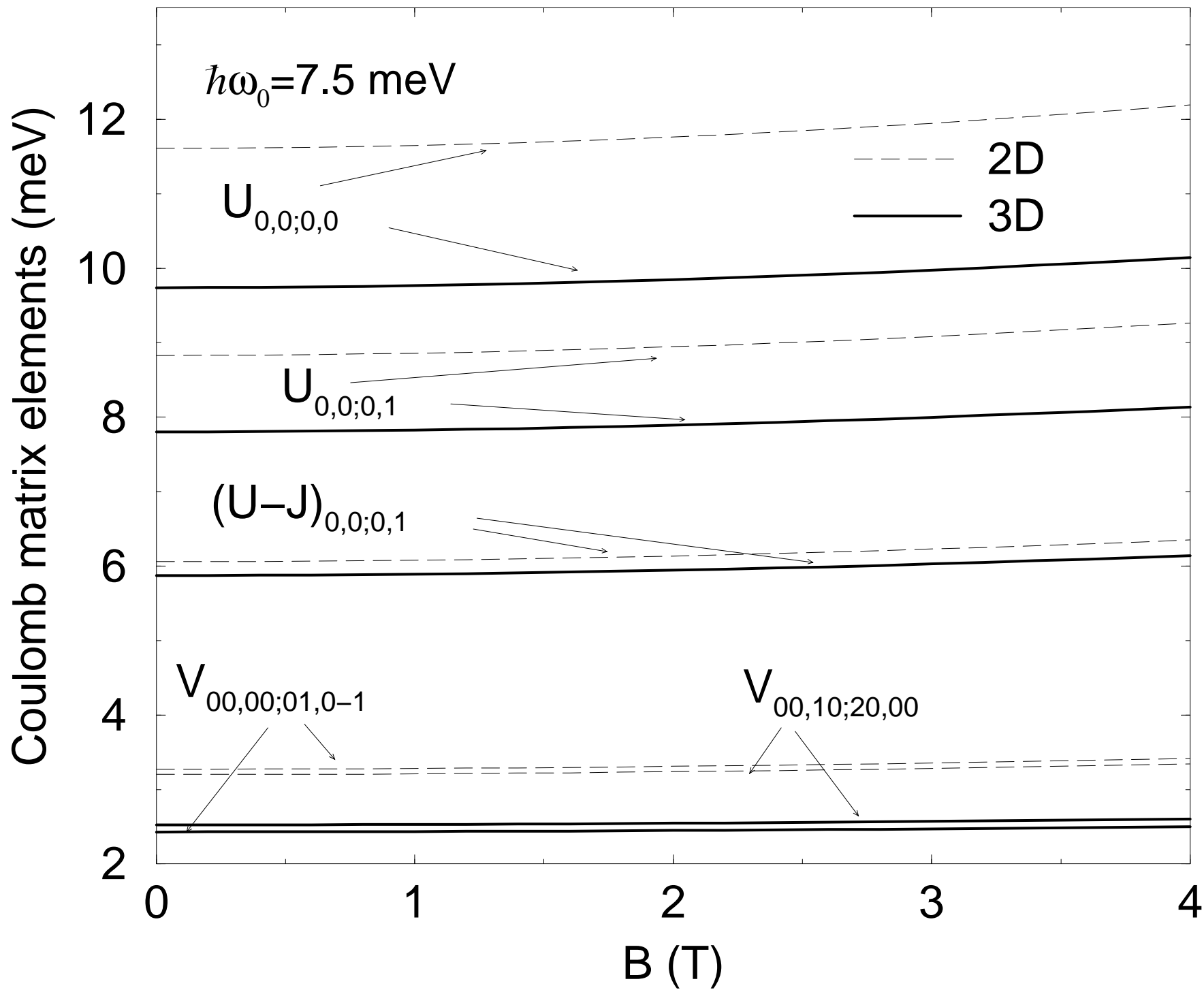
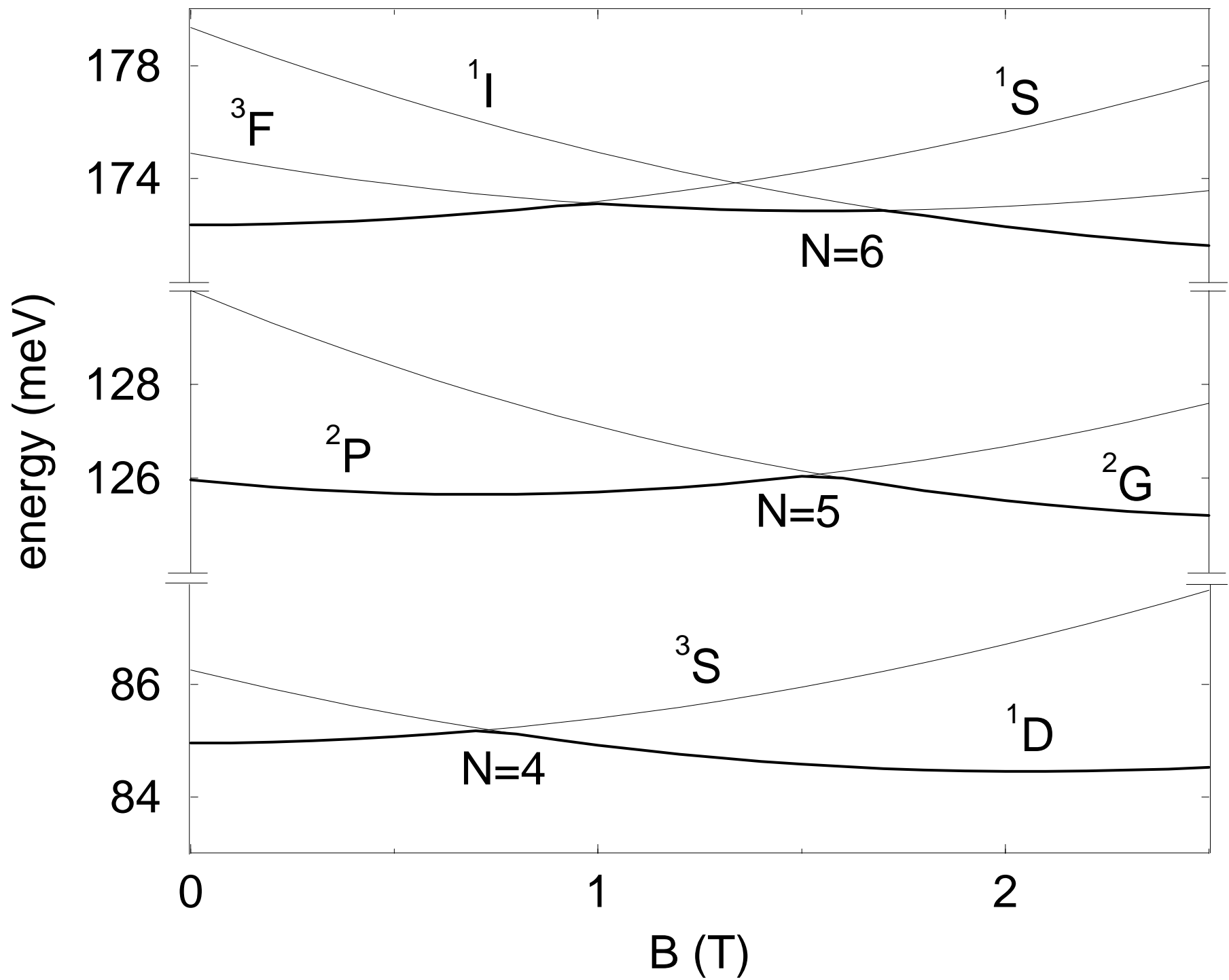
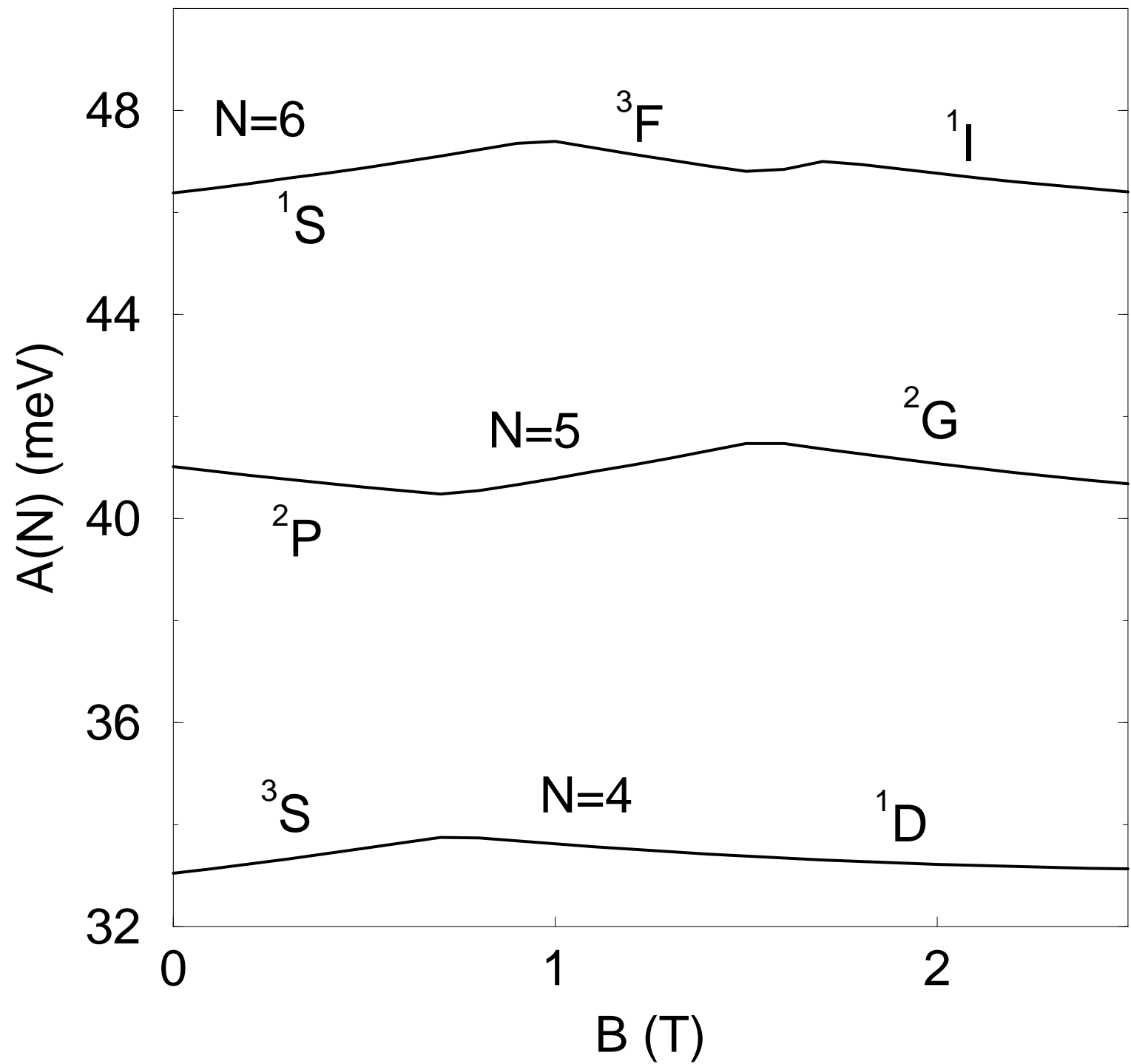
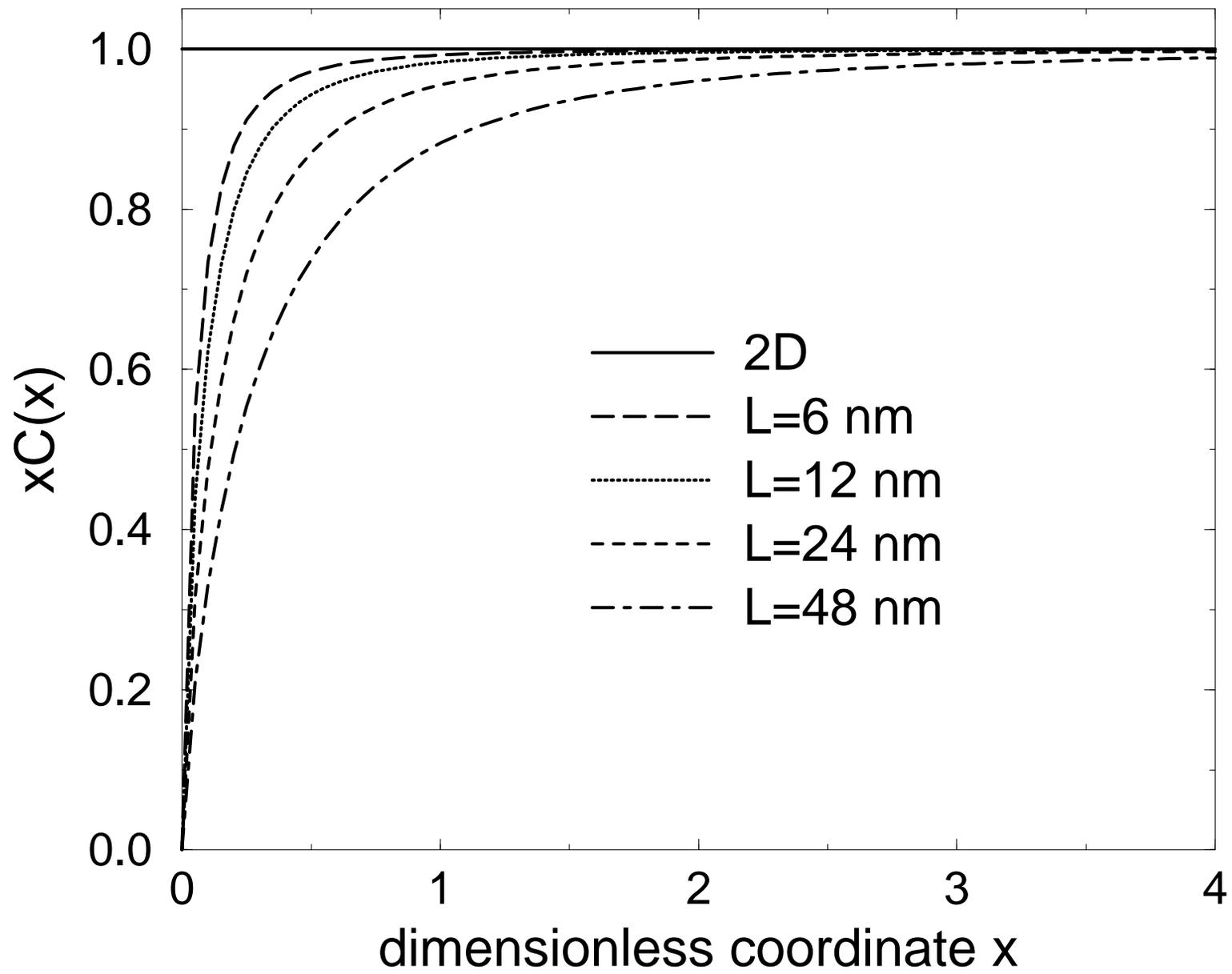
\end{document}